\documentclass[superscriptaddress,twocolumn,amsmath,amssymb,aps,pra,nofootinbib]{revtex4-2}

\usepackage{graphicx}
\graphicspath{{figures/}}
\usepackage{bm}
\usepackage{amsthm}
\usepackage{mathrsfs}
\usepackage[colorlinks=true,citecolor=blue,linkcolor=blue,urlcolor=blue]{hyperref}

\newcommand{\ket}[1]{\lvert #1\rangle}
\newcommand{\bra}[1]{\langle #1\rvert}

\newcommand{\Rea}{\operatorname{Re}}

\newcommand{\dd}{\mathrm{d}}

\begin{document}

\title{Optimal quantum metrology with restricted controls}
\author{Li Sun}
\email{lisun001@cuhk.edu.hk}
\affiliation{Department of Physics, The Chinese University of Hong Kong, Shatin, New Territories, Hong Kong, China}
\author{Haoze Wang}
\affiliation{Department of Physics, The Chinese University of Hong Kong, Shatin, New Territories, Hong Kong, China}
\author{Chong Chen}
\email{chongchenn@gmail.com}
\affiliation{Department of Physics, The Chinese University of Hong Kong, Shatin, New Territories, Hong Kong, China}
\date{\today}

\begin{abstract}
Quantum control can enhance sensing precision, but practical protocols are constrained by probe preparation and the available control resources. Optimizing the sensing protocol within these constraints helps establish what precision is experimentally attainable.  We study quantum metrology from a fixed initial probe state under restricted Hamiltonian controls with bounded amplitudes. Building on Pontryagin's maximum principle, we formulate quantum Fisher information maximization at a prescribed interrogation time for multiple control channels subject to a weighted amplitude constraint. The formulation also accommodates stochastic Hamiltonian evolution and time-local master equations. A geometric description of sensitivity accumulation reveals the mechanisms underlying the optimized protocols. Two spin-$1/2$ examples demonstrate the resulting enhancement: time-dependent control outperforms optimized constant control in transverse-field sensing, while optimized waveforms improve upon CPMG pulses for AC-field sensing under strong colored noise. These findings highlight the importance of optimizing sensing protocols within available control resources to improve experimentally attainable precision.
\end{abstract}

\maketitle

\section{Introduction}
Quantum metrology exploits quantum effects to enhance the precision of parameter estimation~\cite{Giovannetti2004,Giovannetti2006,Giovannetti2011,Pezze2018}. In a typical sensing protocol, the parameter is encoded in the dynamics of a prepared probe and inferred from measurements of its final state~\cite{Degen2017}. Achieving the best possible precision requires optimizing the initial probe state, the encoding dynamics, and the final measurement~\cite{Liu2022, Zhang2022QuanEstimation, Liu2024FullyOptimized}. For a fixed encoding protocol, such as Ramsey interferometry, entanglement and spin squeezing are established resources for enhanced sensitivity~\cite{KitagawaUeda1993,Wineland1994,Ma2011}. For a given parameter-dependent final state, the quantum Fisher information (QFI) determines the quantum Cram\'er--Rao bound on estimation precision, accounting for optimization over all measurements~\cite{Helstrom1976,BraunsteinCaves1994,Paris2009}. Quantum control provides a complementary route to higher precision
by shaping the encoding dynamics. It can improve the accumulation
of parameter sensitivity when the Hamiltonian does not commute with
its parameter derivative~\cite{YuanFung2015,PangJordan2017,Hou2019}
and balance signal accumulation against noise-induced
decoherence~\cite{Poggiali2018,HernandezGomez2024,Timms2025}.
Identifying optimal controls for such noncommuting dynamics and noisy environment is therefore central to establishing and attaining the precision limits of quantum sensing.

Experimental resources constrain both probe preparation and dynamical control~\cite{Pezze2018}. These tasks are most readily implemented in simple probes, such as single-spin sensors, and become more demanding in many-body systems. For example, preparing an $N$-body Greenberger--Horne--Zeilinger state requires coherent entangling operations across all $N$ particles, with control errors and decoherence limiting the achievable fidelity~\cite{Omran2019, Bao2024}. The initial probe state may consequently be fixed by the available preparation procedure, for example as the ground state of a known Hamiltonian. Control is likewise typically limited to experimentally accessible Hamiltonian terms with bounded amplitudes, such as external-field couplings~\cite{Koch2022,YangEtAl2022}. Decoherence further restricts the useful interrogation time~\cite{Huelga1997,Escher2011}, motivating optimization over a finite sensing duration.  Based on these constraints, it is natural to ask: what is the maximum QFI attainable at a prescribed interrogation time $T$ from a fixed initial state under bounded, restricted control?

Quantum optimal control theory provides a framework for optimizing sensing protocols under experimental constraints~\cite{Koch2022}. Numerical optimization~\cite{Khaneja2005,LiuYuan2017,Liu2022}, Pontryagin's maximum principle~\cite{Boscain2021,LinMaSels2021,Lin2022}, and metrological variational methods~\cite{YangEtAl2022} offer complementary approaches. Numerical optimization has also been applied to pulse
sequences and continuous control waveforms for sensing in noisy
environments~\cite{Poggiali2018,HernandezGomez2024,Timms2025}. In particular, Pontryagin’s maximum principle (PMP) enables efficient evaluation of the QFI gradient with respect to the control fields and has been applied to optimal sensing with interacting spin ensembles under bounded controls, both with and without decoherence~\cite{LinMaSels2021,Lin2022}. Yang et al.~\cite{YangEtAl2022} also developed a variational principle for jointly optimizing the initial probe state and controls within a prescribed operator subspace. Building on these approaches, we investigate how restrictions on the available control operators and their amplitudes affect the sensing precision.

In this work, we apply PMP to optimize quantum sensing under restricted controls. Building on the PMP-based formulations of Refs.~\cite{LinMaSels2021,Lin2022}, the QFI optimization from a prescribed initial probe state with multiple Hamiltonian control channels subject to a weighted amplitude constraint is investigated for unitary evolution, noise-averaged stochastic trajectories, and non-unitary dynamics with a time-local master equation.  The optimal sensing protocols under these constraints are obtained and the physical mechanisms responsible for their performance are explained.  Two different sensing schemes based on a spin-$1/2$ sensor are used to demonstrate the power of this variational method. One is transverse field sensing, in which the intrinsic Hamiltonian cannot be made to commute with the sensing term even with the available controls.  Results show that constrained sensing can require nontrivial control protocols even for a single spin. The second example is AC-field sensing under colored noise. Under strong noise, we show that a complex control waveform yields a substantially larger terminal QFI than CPMG pulses with more effectively preserving the coherence of the quantum sensor and without substantially compromising AC-signal accumulation. The sensitivity trajectories provide a geometric interpretation of how the optimized control balances dephasing suppression and parameter-sensitivity accumulation. Together, these examples demonstrate how optimal control can improve sensing performance.

The paper is organized as follows. In Sec.~\ref{sec:ParameterEstimation}, we introduce the parameter-estimation problem with a fixed initial probe state and bounded Hamiltonian controls. In Sec.~\ref{sec:PMP}, we derive the state--costate equations and control gradient using Pontryagin's maximum principle and present the projected gradient ascent algorithm. In Sec.~\ref{sec:Generalization}, we extend the formulation to stochastic Hamiltonian evolution and time-local master equations. In Sec.~\ref{sec:QFIAccumulation},  a geometric description of QFI accumulation in terms of the sensitivity-tangent speed,  alignment factor, and accumulation phase is introduced to help explain the advantages of optimized control. In Sec.~\ref{sec:NuermicalResults}, we apply the framework to transverse-field sensing and AC-field sensing under colored noise, highlighting the performance and physical mechanisms of the optimized protocols. Section~\ref{sec:Conclusions} summarizes the results and discusses future directions.

\section{parameter estimation with restricted control}\label{sec:ParameterEstimation}
We consider single-parameter estimation with a fixed initial probe state. Here, we take a pure initial state and unitary evolution as a specific example. The probe is prepared in a pure state $\ket{\psi_0}$, independent of the unknown parameter $\theta$, and evolves under the time-dependent Hamiltonian 
\begin{equation}\label{eq:Hamiltonian}
    H(\theta,t)=H_{\theta}(t)+H_c(t),
\end{equation}
where $H_{\theta}(t)$ is the uncontrolled Hamiltonian, with $\theta$ denoting the parameter to be estimated, and $H_c(t)$ is a parameter-independent control Hamiltonian. Here, we set
\begin{equation}
    H_c(t)=\sum_a u_a(t) A_a,
\end{equation}
where $\{A_a\}$ is a fixed set of accessible Hermitian operators and $u_a(t)$ is the control amplitude. At a fixed sensing time $T$ due to the irreversible decoherence, the final state reads $|\psi_T\rangle=\mathcal{T}\exp[-i \int^{T}_0 dt H(\theta,t) ]\ket{\psi_0}$, where $\mathcal{T}$ denotes the time ordering. For an unbiased estimation with a given readout observable $O$, the sensitivity is determined by the signal-to-noise ratio
\begin{equation}
    \text{SNR}_{O}=\frac{| \langle\phi_T|O | \psi_T\rangle+\langle\psi_T|O | \phi_T\rangle|^2}{\langle\psi_T|O^2 | \psi_T\rangle-\langle \psi_T| O|\psi_T\rangle^2},
\end{equation}
where $|\phi_T\rangle\equiv |\partial_{\theta}\psi_T\rangle$ is the parameter derivative. This sensitivity satisfies $\mathrm{SNR} \leq F_\theta$, where the pure-state quantum Fisher information (QFI) is
\begin{equation}\label{eq}
F_\theta=4[\langle \phi_T|\phi_T\rangle-\left|\langle \psi_T|\phi_T\rangle\right|^2].
\end{equation}
The bound is saturated by an optimal readout observable, yielding $\max_O\mathrm{SNR}_O=F_\theta$. Note that the parameter derivative $\ket{\phi_t}$ also obeys a time-local equation of motion
\begin{equation}\label{eq:evolution-phi}
i\partial_t\ket{\phi_t}=H(\theta,t)\ket{\phi_t}
+ K(t)\ket{\psi_t},
\end{equation}
where $\ket{\phi_0}=0$ and $K(t)  \equiv \partial_\theta H_\theta(t)$. 

In addition to restricting the available control operators to $\{A_a\}$, we impose an amplitude constraint reflecting finite experimental resources:
\begin{equation}\label{eq:Constrain}
  C(t)= \sum_{a} q_{a} u_{a}^2(t)-E^2 \le 0,
\end{equation}
where $q_a \geq 1$ weights the contribution of each control field and $E$ sets the overall amplitude bound. This constrain complicates the optimization but ensures that the resulting protocols are relevant to experiments. From this constraint, one can find that a larger weight $q_a$ imposes a tighter amplitude bound on the corresponding control channel. Following restricted-control constructions~\cite{WangEtAl2015}, we seek the admissible control protocol that maximizes the final-state QFI.

\section{Pontryagin maximum principle}\label{sec:PMP}
To optimize the final QFI under the prescribed control constraints, we apply  PMP~\cite{Boscain2021,LinMaSels2021,Lin2022}. It yields coupled state--costate equations and an explicit QFI gradient with respect to the controls.  By combining the probe state and its parameter derivative into an augmented state $\bm z=(\ket{\psi_t},\ket{\phi_t})^{\mathsf T}$, the Schr\"{o}dinger equation and Eq.~\eqref{eq:evolution-phi} then give
\begin{equation}
    i\dot{\bm z}=\mathbb L\bm z,~~  \mathbb L=\left(\begin{matrix}H(t) &0\\K(t) &H(t) \end{matrix}\right),
\end{equation}
where the initial condition reads $\bm z(0)=(\ket{\psi_0},0)^{\mathsf T}$. To maximize the final QFI with restricted control, we introduce the constrained action
\begin{equation}\label{eq:joint_action}
\mathcal S=F_\theta+\int_0^T\dd t\,\Bigl[
2\Rea\,\bm\chi^\dagger(i\dot{\bm z}-\mathbb L\bm z)
-\lambda_t C(t)
\Bigr],
\end{equation}
where $\bm\chi=(\ket{\chi_\psi},\ket{\chi_\phi})^{\mathsf T}$ is the costate enforcing the dynamics, and $\lambda_t\ge0$ is the multiplier associated with the amplitude constraint given in Eq.~\eqref{eq:Constrain}. Using this action, the constrained QFI optimization can be expressed as
\begin{equation}
    \max_{u_t, C(t)\le 0} F_\theta[u_a(t)] = \max_{u,\bm{z}} \inf_{\bm{\chi},\lambda_t \ge 0} \mathcal{S}[u_a(t),\lambda_t, \bm{z}, \bm{\chi}],
\end{equation}
where the inner infimum excludes infeasible trajectories and equals the final QFI on feasible ones. The propagator-based formulation of Ref.~\cite{YangEtAl2022} uses a time-integrated objective and jointly optimizes the initial state and encoding dynamics. For a prescribed initial probe state, our objective depends only on the terminal augmented state $\bm z(T)$, which has the same terminal-cost structure as the one used in Ref.~\cite{LinMaSels2021,Lin2022}. Moreover, when the measured observable $O$ is fixed experimentally, $F_\theta$ can instead be replaced by $\mathrm{SNR}_O$ to optimize the precision attainable with that observable.

Stationarity of the action with respect to the costate $\bm{\chi}$ and state $\bm z$ yields the coupled equations
 \begin{align} \label{eq:joint_system}
  i\dot{\bm z}&= \mathbb L\bm{z}, \quad i\dot{\bm\chi}= \mathbb L^{\dagger}\bm\chi. 
 \end{align}
In addition, the variation of the action with respect to terminal state $z(T)$ yields the boundary condition 
\begin{equation}
     \bm\chi(T)= 4i(c\ket{\phi_T},  c^{\ast} |\psi_T\rangle -|\phi_T\rangle)^T,
\end{equation}
where $c=\langle \phi_T|\psi_T\rangle$. Together with the initial condition $\bm z(0)=(\ket{\psi_0},0)^{\mathsf T}$, the state $\bm{z}(t)$ and costate $\chi(t)$ can be solved forward and backward, respectively. Furthermore, control stationarity and minimization over nonnegative multipliers yield the Karush--Kuhn--Tucker (KKT) conditions~\cite{BoydVandenberghe2004, NocedalWright2006}
\begin{equation}\label{eq:KKT}
G_{a}(t)-\lambda_t\frac{\partial C(t)}{\partial u_{a}(t)}=0,
\qquad
C(t)\le0,\quad \lambda_tC(t)=0,
\end{equation}
where $G_{a}(t)\equiv \frac{\delta F_\theta[u]}{\delta u_{a}(t)}$ is the QFI gradient. The gradient is given by
\begin{equation}\label{eq:Gradient-QFI}
G_{a}(t)
=-2\Rea\left[
\bra{\chi_\psi}A_{a}\ket{\psi_t}
+\bra{\chi_\phi}A_{a}\ket{\phi_t}\right].
\end{equation}
Complementary slackness, $\lambda_tC(t)=0$, implies that the multiplier vanishes whenever the amplitude bound is inactive. These equations provide necessary conditions for optimality and do not, by themselves, guarantee global optimality.

\subsection{projected gradient ascent}\label{projected gradient ascent}

For fixed control weights $\{q_{a}\}$, the control fields $u_{a}(t)$ can be optimized by projected gradient ascent. At iteration $n$, we propagate the state $\bm z$ forward and the costate $\bm\chi$ backward using the current controls $u_{a}^{(n)}(t)$. The QFI gradient $G_{a}^{(n)}(t)$ in Eq.~\eqref{eq:Gradient-QFI} then defines the trial controls $v_{a}(t)=u_{a}^{(n)}(t)+\eta G_{a}^{(n)}(t)$, where the step size $\eta>0$ is chosen by a line search to increase the final QFI after projection. We enforce the amplitude constraint by projecting the trial controls onto the admissible set:
\[
u_{a}^{(n+1)}(t)=\frac{v_{a}(t)}{1+q_{a}r_t}.
\]
If the trial controls satisfy the amplitude constraint, $r_t=0$. Otherwise, $r_t>0$ is the unique solution of
$\sum_{a} q_{a}v_{a}^2(t)/(1+q_{a}r_t)^2=E^2$. Each projection preserves feasibility and satisfies the KKT
conditions for the projection subproblem. The iterations stop when the projected update is sufficiently small. At a fixed point, the control satisfies the KKT conditions in Eq.~\eqref{eq:KKT}, with $\lambda_t=r_t^\ast/(2\eta^\ast)$.

\section{Mixed state and general time-local dynamics}\label{sec:Generalization}
The preceding analysis can be generalized to mixed initial states and general time-local dynamics. Two complementary descriptions are considered: stochastic
Hamiltonian evolution and deterministic time-local master
equations.  In both cases, the final state is a mixed state and thus the calculation of QFI should be generalized to mixed state case. 

For a parameter-dependent mixed state $\rho_T$, its parameter derivative reads
$\sigma_T=\partial_\theta\rho_T$. The sensitivity associated with
a fixed Hermitian readout observable $O$ is
\begin{equation}
\mathrm{SNR}_O=
\frac{|\operatorname{Tr}(O\sigma_T)|^2}
{\operatorname{Tr}(O^2\rho_T)
-[\operatorname{Tr}(O\rho_T)]^2}.
\end{equation}
According to quantum Cram\'{e}r-Rao bound~\cite{BraunsteinCaves1994}, the optimal readout observable $L$ is determined by the symmetric logarithmic derivative
\begin{equation}\label{eq:SLD}
\frac{1}{2}(L\rho_T+\rho_T L)=\sigma_T.
\end{equation}
It yields the QFI
\begin{equation}
F_\theta=\operatorname{Tr}(\rho_T L^2).
\label{eq:QFI-MS}
\end{equation}
Above derivation reveals that QFI is completely determined by of $\rho_T$ and $\sigma_T$. The first variation over them yields
\begin{equation}
\delta F_\theta=
2\operatorname{Tr}(L\,\delta\sigma_T)
-\operatorname{Tr}(L^2\delta\rho_T).
\label{eq:QFI-variation}
\end{equation}

\subsection{Stochastic Hamiltonian evolution} \label{subsec:Generalization-SHE}

We first consider stochastic unitary trajectories generated by a noisy
Hamiltonian~\cite{Chenu2017},
\begin{equation}
H_\xi(\theta,t)=H_{\theta,\xi}(t)+\sum_a u_a(t)A_a,
\end{equation}
where $\xi$ labels the noise dependent and $u_a(t)$ is the control field amplitude. We assume that the noise distribution is independent of the parameter and controls, and the same control protocol is applied to all realizations.

Starting from a fixed, pure state $|\psi_0\rangle$, the state $|\psi_\xi(t)\rangle$ and its parameter derivative $|\phi_\xi(t)\rangle$ obey
\begin{equation}
    i\dot{\bm{z}}_\xi (t)= \left( \begin{array}{cc}
    H_\xi(t) & 0\\
    K(t) & H_\xi(t)\\
    \end{array}
    \right) \bm{z}_\xi(t),
\end{equation}
where $\bm{z}(t)=(|\psi_\xi(t)\rangle, |\phi_\xi(t)\rangle)$. For unobserved noise, the terminal density matrix and its parameter derivative are
\begin{align}
\rho_T&=\mathbb E_\xi
\left[\ket{\psi_\xi(T)}\bra{\psi_\xi(T)}\right],\\
\sigma_T&=\mathbb E_\xi
\left[
\ket{\phi_\xi(T)}\bra{\psi_\xi(T)}
+\mathrm{H.c.}
\right],
\end{align}
where $\mathbb E_\xi$ denotes the noise average. 

Under the stochastic evolution, the constrained action  shown in Eq.~\eqref{eq:joint_action} becomes
\begin{equation}
\mathcal S=F_\theta+\int_0^T\dd t\, \Bigl[
2\Rea\, \mathbb{E}[\bm\chi^\dagger_\xi(i\dot{\bm z}_\xi-\mathbb L_\xi\bm z_\xi)]
-\lambda_t C(t)
\Bigr],
\end{equation}
Variation with respect to the states and costates yields
\begin{equation}
i\dot{\bm z}_\xi=\mathbb L_\xi\bm z_\xi,
\qquad
i\dot{\bm\chi}_\xi=\mathbb L_\xi^\dagger\bm\chi_\xi.
\end{equation}
Furthermore, using the detailed expression of $\rho_T$ and $\sigma_T$ and the first variation of QFI, one can get these variations equations
\begin{align}
        \partial_{\langle \psi_\xi(T)|}F_\theta= & -L^2 |\psi_\xi(T)\rangle +2 L |\phi_\xi(T)\rangle, \\
        \partial_{\langle \phi_\xi(T)|}F_\theta= & 2 L |\psi_\xi(T)\rangle.
\end{align}
Combining this variation with the boundary term from the action then gives
\begin{equation}
\bm\chi_\xi(T)=
i\begin{pmatrix}
L^2\ket{\psi_\xi(T)}-2L\ket{\phi_\xi(T)}\\
-2L\ket{\psi_\xi(T)}
\end{pmatrix},
\end{equation}
where $L$ can be obtained from solving Eq.~\eqref{eq:SLD}.

Under stochastic evolution, the QFI gradient with respect to the control amplitude $u_a(t)$ is
\begin{align}
G_a(t)=-2\operatorname{Re}\mathbb E_\xi\Bigl[
&\bra{\chi_{\psi,\xi}}A_a\ket{\psi_\xi} +\bra{\chi_{\phi,\xi}}A_a\ket{\phi_\xi}
\Bigr].
\end{align}
The control stationarity and complementary-slackness conditions
retain the form given in Eq.~\eqref{eq:KKT}. The gradient can
therefore be used in the same projected gradient ascent procedure.

\subsection{Time-local master equations}

Another generalization is for nonunitary evolution governed by a time-local master equation~\cite{BreuerPetruccione2002},
\begin{equation}
\dot\rho_t=\mathcal L(t)\rho_t,
\end{equation}
where $\mathcal L(t)$ is a linear, trace-preserving and Hermiticity-preserving generator. In finite dimensions, such a generator admits a Lindblad-like representation~\cite{Gorini1976,Lindblad1976,Hall2014},
\begin{equation}
\mathcal L(t)\rho=-i[H(t),\rho]
+\sum_{ij}\gamma_{ij}(t)\Bigl(
O_i\rho O_j^\dagger -\frac12\{O_j^\dagger O_i,\rho\}\Bigr),
\end{equation}
where $\gamma_{ij}(t)$ is a Hermitian rate matrix, $[A,B]\equiv AB-BA$ and $\{A,B\}\equiv AB+BA$ are the commutator and anticommutator of $A$ and $B$, respectively.

As in the unitary case, here we have 
\begin{equation}
    \mathcal{L}(t)=\mathcal{L}_\theta(t)+\mathcal{L}_c (t),
\end{equation}
where 
$ \mathcal{L}_c(t)=\sum_{a} u_a(t) \mathcal{A}_a$ and  $\{\mathcal{A}_a\}$ is the set of available control superoperators. Here, $\{u_a(t)\}$ satisfy the same bound shown in Eq.~\eqref{eq:Constrain}.

As in the pure state case, the parameter derivative $\sigma_t\equiv \partial_\theta \rho$ also satisfies a time-local evolution
\begin{equation}
    \frac{d}{dt}\sigma_t= \mathcal{L}(t) \sigma_t + \mathcal{K}(t) \rho_t,
\end{equation}
where $\mathcal{K}(t)\equiv \partial_\theta \mathcal{L}(t)$. By taking $\bm{z}=(\rho_t, \sigma_t)$ as the augmented state, the constrained action then reads
\begin{equation}
\mathcal S=F_\theta+\int_0^T\dd t\,\Bigl[
\text{Tr}[\bm\chi (\dot{\bm z}-\mathbb L\bm z)]
-\lambda_t C(t)
\Bigr],
\end{equation}
where $\bm{\chi}$ is the corresponding costate and $\mathbb L=\bigl(\begin{smallmatrix} \mathcal{L}&0\\ \mathcal{K}& \mathcal{L}\end{smallmatrix}\bigr)$.  Using the same variational method, one can get the evolution of state and costate, given by 
\begin{equation}
    \dot{\bm z}=\mathbb{L} \bm{z},~~ \dot{\bm \chi}=-\mathbb{L}^{\dagger} \bm{\chi}
\end{equation}
with the boundary conditions $\bm{z}(0)=(\rho_0,0)$ and $\bm{\chi}(T)=(L^2, -2L)$. Here $\text{Tr}[\bm{z} \mathbb{L}^{\dagger} \bm{\chi}]=\text{Tr}[\bm{\chi} \mathbb{L} \bm{z}]$. In addition, the QFI gradient defined in Eq.~\eqref{eq:KKT} becomes
\begin{equation}
    G_a(t)=-\text{Tr}[\bm{\chi} \left( \begin{matrix}  \mathcal{A}_a & 0 \\ 0 &\mathcal{A}_a \end{matrix} \right) \bm{z}].
\end{equation}
The same forward--backward propagation and projected gradient ascent algorithm can thus be used to optimize the final QFI under controls.

\section{QFI accumulation}\label{sec:QFIAccumulation}
To characterize the accumulation of QFI during evolution, we introduce a purification amplitude $W=\sqrt{\rho}U$, where $U$ is unitary to be determined and $\rho=WW^\dagger$. By properly choosing $U$, the purification can be transported horizontally along $\theta$, yields the horizontal lift of the parameter derivative $\partial_\theta\rho$ at $W$~\cite{Uhlmann1986,DittmannUhlmann1999}, given by
\begin{equation}
    \eta=\partial_\theta W=\frac12LW,
\end{equation}
where $L$ is the symmetric logarithmic derivative, defined in Eq. \eqref{eq:SLD}.  The norm of $\eta$ determines the QFI through $F_\theta=4\operatorname{Tr}[\eta^\dagger\eta]$. Due to the existence of $U$, the ordinary derivative $\dot\eta$ includes motion associated with the arbitrary purification gauge. For a specified coherent Hamiltonian $H(t)$, we decompose the amplitude velocity as
\begin{equation}
\dot W=\left(-iH+\frac12L_D\right)W+WA_t,
\end{equation}
where $A_t^\dagger=-A_t$ describes purification gauge motion. The Hermitian operator $L_D$ is determined by
\begin{equation}
\dot\rho+i[H,\rho] =\frac12(L_D\rho+\rho L_D).
\end{equation}

To remove gauge motion and homogeneous Hamiltonian transport, we define
\begin{equation}
D_t\eta=\dot\eta+iH\eta-\eta A_t.
\end{equation}
Under a parameter-independent gauge transformation $W\mapsto WV(t)$, this derivative transforms covariantly, $D_t\eta\mapsto(D_t\eta)V$. Its Hilbert-Schmidt norm and
its overlap with $\eta$ are consequently gauge invariant. Substituting $\eta=LW/2$ gives
\begin{equation}
D_t\eta= \frac12\left(\dot L+i[H,L]+\frac12 L L_D\right)W.
\label{eq:Dteta}
\end{equation}

Note that the removed transport terms have zero real overlap with $\eta$. Consequently, one can find that
\begin{equation}\label{eq:QFIAccumulation}
\frac{d\sqrt{F_\theta}}{dt}
=2 J_{F} \kappa \cos\beta,
\qquad
\left|\frac{d\sqrt{F_\theta}}{dt}\right|\le 2 J_F \kappa,
\end{equation}
where $J_F=|D_t\eta|$ and $\kappa=|\operatorname{Tr}(\eta^\dagger D_t\eta)|
/|\eta| |D_t\eta|$, $\beta=\rm{arg}[\operatorname{Tr}(\eta^\dagger D_t\eta)]$, and $|\cdot|$ denotes the Hermite-Schmidt norm. Here, $J_F$ is the sensitivity-tangent speed, $\kappa\in[0,1]$ is the alignment factor, and $\beta$ is the accumulation phase. The product $J_F\kappa$ is the magnitude of the projection of $D_t\eta$ onto the complex line spanned by $\eta$, which we call the parallel sensitivity speed. Its real component, $J_F\kappa\cos\beta$, determines sensitivity growth. The individual quantities depend on the specified transport convention, whereas their combination in Eq.~\eqref{eq:QFIAccumulation} gives the sensitivity growth rate.

For pure-state unitary evolution, the construction can be introduced directly. Locally at $\theta=\theta_0$, choose $W=e^{-(\theta-\theta_0) \langle \psi_t| \phi_t\rangle_{\theta=\theta_0}} |\psi_t\rangle \langle a|$ with $|a\rangle$ being a fixed, normalized auxiliary state. The horizontal parameter tangent then reads $\eta=|\phi_\perp\rangle\langle a|$ with $|\phi_\perp\rangle= |\phi_t \rangle-|\psi_t\rangle\langle\psi_t|\phi_t \rangle$. Defining the centered operator $\delta K=K(t)-\langle\psi_t|K(t)|\psi_t\rangle$, we obtain 
\begin{equation}
D_t\eta=-i \delta K(t)W.
\end{equation}
Consequently, the sensitivity-tangent speed reads
\begin{align}
    J_F = & \sqrt{\langle \psi_t|\delta K^2(t) |\psi_t\rangle}.
\end{align}
The alignment factor and the accumulation phase are
\begin{equation}
   \kappa = \frac{| \langle \phi_\perp|  K(t) |\psi_t\rangle|}{J_F |\phi_\perp| }, ~~ \beta={\rm Arg} [-i \langle \phi_\perp| K(t) |\psi_t\rangle].
\end{equation}

\begin{figure*}[t]
\centering
\includegraphics[width=1.0\textwidth]{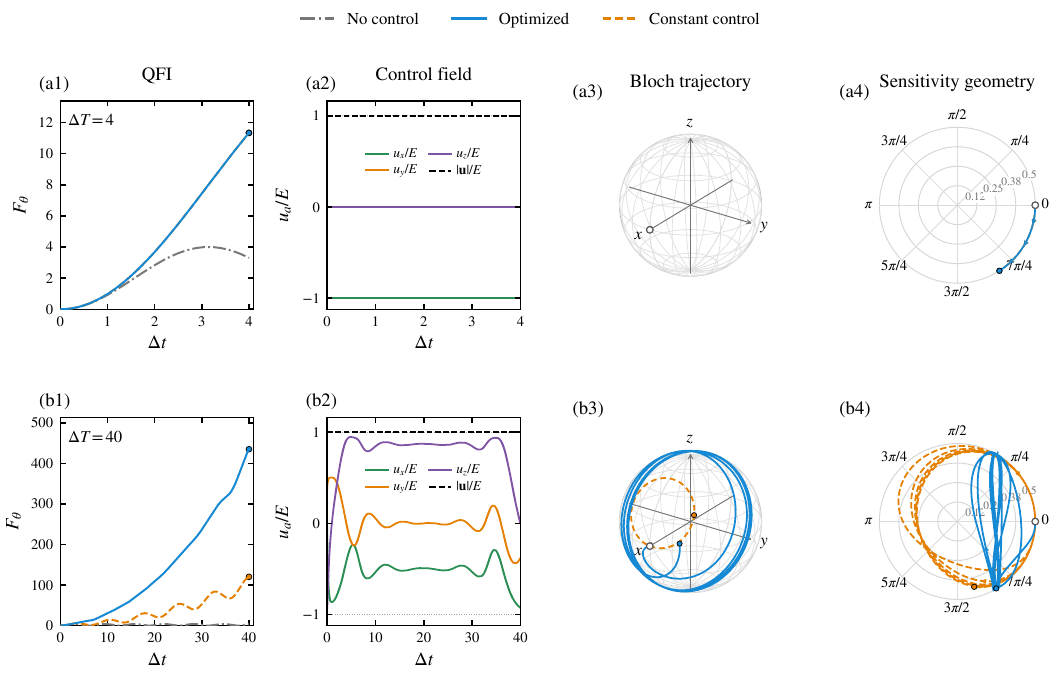}
    \caption{Optimal control for transverse-field sensing with a spin-$1/2$ sensor under restricted controls $|\bm{u}|\le E$.  The interrogation times are $\Delta T=4.0$ (top row) and $\Delta T=40.0$ (bottom row), representing short- and long-time sensing, respectively. (a1, b1) QFI under the optimized protocol, compared with uncontrolled case and the optimized constant control $\bm{u}^{\ast}$. (a2, b2) the change of optimal control field with time.  (a3, b3) Bloch-sphere trajectory for different cases.  (a4, b4) Polar plots with the parallel sensitivity speed $J_F \kappa$ and accumulation phase $\beta$ as the radial and angular coordinates, respectively. Parameters: $\Delta=1.0$ and $E=0.5\Delta$, with initial state $|+x\rangle$ and operating point $\theta_0=0$.}
    \label{fig:qubit_main}
\end{figure*}

\section{Numerical results}\label{sec:NuermicalResults}

In this section, the variational principle is illustrated by two examples. One is the transverse-field sensing with a spin-$1/2$ sensor, where the Hamiltonian and its parameter derivative correspond to orthogonal directions. Another example is the AC field sensing with a spin-$1/2$ sensor subject to colored noise, modeling the nuclear spin detection with a central spin~\cite{Maze2008,Titum2021, Timms2025}. 

\subsection{Transverse field sensing}
For the transverse-field sensing, we consider a spin-$1/2$ sensor with an intrinsic Hamiltonian along the $x$ direction and an unknown field along the $z$-direction. The Hamiltonian then reads
\begin{equation}
H_\theta=\Delta S_x+\theta S_z,\qquad H_c=\sum_{a=x,y,z}u_a(t)S_a,
\label{eq:main_qubit_model}
\end{equation}
where $S_a=\sigma_{a}/2$ are the spin-$1/2$ operators. The  control amplitudes satisfy 
\[u_x^2+u_y^2+u_z^2\leq E^2,\]
where we use equal weights $q_a=1$. Here, we set $E<\Delta$, which describes the transverse field sensing of a well-split spin-1/2 sensor under control fields of finite amplitude. The initial state is fixed to $\ket{\psi_0}=\ket{+_x}$ and we set $\theta_0=0$. The non-commutativity between intrinsic Hamiltonian and its parameter derivative hints that the optimal control can enhance the sensitivity.  

For a spin-1/2, the state at time $t$ can be fully described by the Bloch vector $\bm r(t)$ as $\rho(t)=\frac{\sigma_0}{2}+\frac{\bm{r}(t)\cdot \bm{\sigma}}{2}$. Its parameter derivative is denoted by $\bm s(t)=\partial_\theta\bm r_\theta(t)$. For pure states, the QFI has a simple expression 
\begin{equation}
    F_\theta=|\bm s|^2.
\end{equation}
In addition, the equations of motion for $\bm{r}$ and $\bm{s}$ can be represented as
\begin{equation}
    \dot{\bm r}= \bm h\times\bm r, ~~\dot{\bm s}= \bm h\times\bm s+\bm e_z\times\bm r,
\end{equation}
where $\bm{h}=(\Delta+ u_x(t), u_y(t), u_z(t))$ and $\bm{e}_i$ denotes the normalized vector along $i$-direction. These equations can be solved by using the initial conditions $\bm{r}(0)=(1,0,0)$ and $\bm{s}(0)=\bm{0}$. From these equations, the sensitivity-tangent speed, alignment factor, and  accumulation phase defined in Eq.~\eqref{eq:QFIAccumulation} are obtained as
\begin{align}
J_F =\frac{1}{2}\sqrt{1-r^2_z},~~~ \kappa=1, ~~
\cos\beta&=\frac{\bm s\cdot(\bm e_z\times\bm r)}{|\bm s|\sqrt{1-r^2_z}},
\label{eq:spin_response_angle}
\end{align}
which illustrates how sensitivity accumulation depends on these factors.

For the uncontrolled case, we have 
\[\bm r(t)=(1,~0,~0),~~\bm s(t)=\frac{1}{\Delta}(0, ~\sin \Delta t, ~1- \cos \Delta t),\] 
leading to $F_\theta(t)={4\sin^2(\Delta t/2)}/{\Delta^2}$. The sensitivity-tangent speed, alignment factor, and accumulation phase are given by
\[J_F=\frac{1}{2},~~\kappa=1,~~ \cos\beta= \frac{\sin \Delta t}{2|\sin \Delta t/2|}.\]
One can find that although parallel sensitivity speed $J_F \kappa =1/2$, the non-zero accumulation phase deteriorates the accumulation of the QFI leading to QFI that oscillates in time. 

For the controlled case with fixed initial state and evolution time $T$, we can only obtain the time-dependent controls numerically. A good choice is the constant control $\bm{u}(t)=\bm{u}^{\ast}$, which can be obtained by optimizing the QFI
\begin{equation}
    \bm{u}^{\ast} =\operatorname{arg\,max}_{\bm{u}:|\bm{u}|\le E} |\bm{s}(T)|^2.
\end{equation}
Here $\bm{s}(T)$ has an analytical solution
\begin{equation*}
    \bm{s}(T)= R_{\bm{e}_h} (\omega T)[\bm{g} \times \bm{r}(0)], 
\end{equation*}
where $R_{\bm{e}_h}(\omega T)$ denotes a $\omega T$ rotation along the axis $\bm{e}_h= \bm{h}/|\bm{h}|$ with $\omega=|\bm{h}|$ and
\begin{equation*}
    \bm{g}= (T-\frac{\sin \omega T}{\omega})\frac{u_z}{\omega} \bm{e}_h+ \frac{\sin \omega T}{\omega} \bm{e}_z -\frac{1-\cos \omega T}{\omega} (\bm{e}_h \times \bm{e}_z). 
\end{equation*}

The optimal constant control field depends on the value of $\Delta T$. At short time $ (\Delta-E) T \lesssim \pi$, numerical optimization recovers the constant control 
\begin{equation}
    u^{\ast}(T)=(-E, 0,0).
\end{equation}
It leads to 
\[F_\theta= 2 \frac{1-\cos(\Delta-E) t}{(\Delta-E)^2}.\]
In this case, the state remains in its initial state as shown in Fig.~\ref{fig:qubit_main} (a3) with a parallel sensitivity speed $J_F \kappa =\frac12$ and a periodically varying accumulation phase with 
\[\cos \beta= \frac{\sin (\Delta-E)t}{2|\sin \frac{(\Delta-E)}{2}t|}.\] 
These results are shown in Fig.~\ref{fig:qubit_main} (a1--a4). 

In contrast, at long times $\Delta T\gg 1$, we have $F_\theta \approx  T^2 u^2_z (u^2_y+u^2_z)/\omega^4+\mathcal{O}(T)$. Maximizing the leading coefficient under $|\bm u|^2\le E^2$ yields 
\begin{equation}\label{eq:constC-LT}
    \bm{u}^{\ast}\approx E(-\frac{E}{\Delta}, 0, \pm \sqrt{1-\frac{E^2}{\Delta^2}}),
\end{equation}
a control field in the $xz$ plane.  For this protocol, the sensitivity-tangent speed and alignment
factor are
\[J_F=\frac12\sqrt{1-\frac{E^2\omega^2}{\Delta^4}
[1-\cos(\omega t)]^2},\qquad \kappa=1.\]
At long times, the accumulation phase satisfies
\[
\cos\beta\approx\frac{1}{2J_F} [\frac{E^2}{\Delta^2} +\frac{\omega^2}{\Delta^2}\cos(\omega t)],
\]
where $\omega=\sqrt{\Delta^2-E^2}$. The resulting growth rate, $\frac{d\sqrt{F_\theta}}{dt}\approx  \frac{E^2}{\Delta^2}+\frac{\omega^2}{\Delta^2} \cos \omega t $, oscillates between $2(E/\Delta)^2-1$ and $1$, with a period average $E^2/\Delta^2$. This gives $F_\theta \approx \frac{E^4}{\Delta^4} T^2 +\mathcal{O}(T)$.  The QFI, the Bloch-sphere trajectory, and the corresponding polar trajectory of $J_F$ and $\beta$ are shown in panels (b1--b4).

Beyond constant control, the control is optimized numerically using those variational equations shown in Sec.~\ref{sec:PMP}. Results are shown in Fig.~\ref{fig:qubit_main} for $E=0.5 \Delta$. At short time $\Delta T=4.0$, one can find that the optimal control is just the constant control with $\bm{u}(t)=(-E, 0, 0)$, which causes a stationary Bloch vector with a fast QFI growth rate $\frac{d\sqrt{F_\theta}}{dt}= \cos \frac{\Delta-E}{2} t$ when $t\le T$. At long interrogation times $\Delta T=40.0$, a nontrivial control protocol emerges and yields a larger QFI than constant control, as shown in Figs.~\ref{fig:qubit_main}(b1--b4).

The Bloch-sphere trajectory in panel (b3) reveals the underlying mechanism. During an initial preparation stage, the control steers the probe toward a more sensitive state whose Bloch vector is perpendicular to the effective field $\bm h=\Delta \bm{e}_x+\bm{u}^{\ast}$ associated with the constant control $\bm u^\ast$ shown in Eq.~\eqref{eq:constC-LT}. The subsequent evolution approximately follows this constant-control Hamiltonian, tracing a great circle around the $\bm{e}_h$ axis. This initial steering improves the probe’s sensitivity during the remaining interrogation time. This behavior is reflected in the trajectory of $J_F \kappa$ and $\beta$ shown in (b4), which illustrates the tradeoff between immediate sensitivity growth and the preparation for subsequent sensing. Compared with constant control (cyan curve), the optimized protocol (red curve) initially sacrifices the growth rate $d\sqrt{F_\theta}/dt=2J_F \kappa \cos\beta$ to achieve a larger QFI growth rate during the subsequent evolution. Furthermore, a final steering stage provides an additional increase in QFI by driving the evolution away from the great circle on the Bloch sphere, also verified in the trajectory of $J_F \kappa$ and $\beta$ shown in (b4). 

Assuming that the intermediate, nearly constant-control stage dominates the interrogation time and that the state is steered onto a great circle about $\bm h$, we estimate 
\begin{equation}
    F_\theta(T)\approx(E/\Delta)^2T^2+\mathcal{O}(T).
\end{equation}
This contrasts sharply with the result $F_\theta(T)\approx(E/\Delta)^4T^2+\mathcal{O}(T)$ obtained by applying the constant control $\bm u^\ast$ from the same initial
state. The improvement over constant control demonstrates the effectiveness of the variational approach in enhancing the QFI for transverse-field sensing under restricted controls.

\subsection{AC field sensing under colored noise}

\begin{figure*}[t]
\centering
\includegraphics[width=1.0\textwidth]{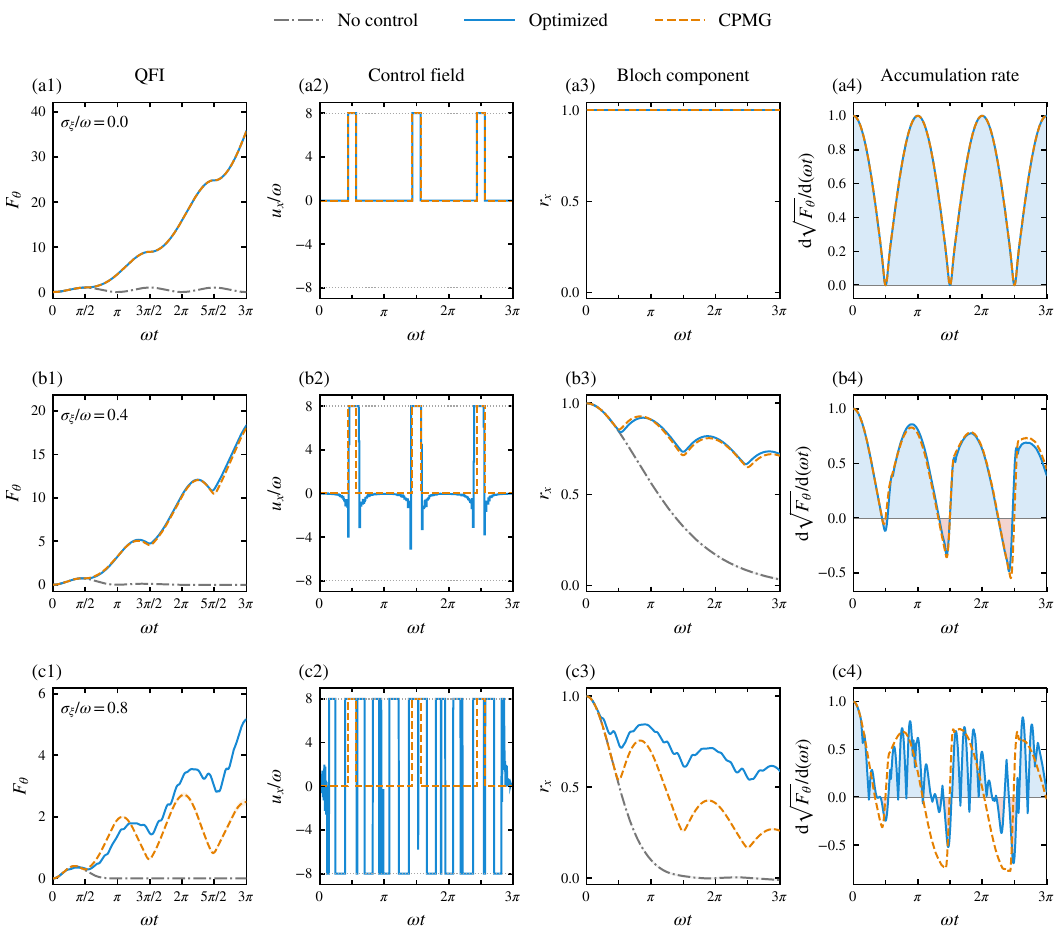}
\caption{Optimized controls for AC-field sensing under colored noise at a fixed interrogation time. The top, middle, and bottom rows show the noise-free, weak-noise, and strong-noise cases, respectively.  (a1--c1)  QFI evolution under uncontrolled field (gray dashed line), optimized control (blue solid line), and  CPMG pulses (orange dashed line). (a2--c2): Optimized control $u_x(t)$ and finite-width CPMG pulses. The control is applied only along the $x$ axis, with $u_y=u_z=0$. (a3--c3): Evolution of $r_x(t)\equiv \langle \sigma_x(t) \rangle$, with $\langle \sigma_y(t) \rangle=\langle \sigma_z(t) \rangle=0$. (a4--c4): evolution of the sensitivity growth $v_F=\frac{d \sqrt{F_{\theta}}}{ d (\omega t)} $. Parameters are: $\omega=1.0$, $\omega \tau_c =\pi$, and $\omega T=3\pi$, with initial state $|+x\rangle$ and operating point $\theta_0=0$. The results are obtained by averaging over $10240$ noise trajectories.}
\label{fig:spin_ac}
\end{figure*}

Another example we consider is a spin-$1/2$ sensor coupled to a weak AC field and subject to colored noise~\cite{Poggiali2018,HernandezGomez2024,Timms2025}. The total Hamiltonian reads $H_{\xi}(\theta, t)=H_{\theta,\xi}(t)+H_c(t)$, with
\begin{equation}\label{eq:spin_ac_model}
H_{\theta, \xi}=[\theta\cos \omega t+\xi(t)]S_z,~ H_c=u_x(t)S_x+u_y(t)S_y,
\end{equation}
where $S_j=\sigma_j/2$ denotes the spin-$1/2$ operator along axis $j$, $\theta$ is the unknown $AC$ field amplitude, and $\omega$ is the known signal frequency. The control fields satisfy
\begin{equation}
     u_x^2+u_y^2\leq E^2,
\end{equation}
corresponding to $q_x=q_y=1$ and $q_z\to\infty$. Here, the noise field $\xi(t)$ is assumed to be a stationary, zero-mean Ornstein--Uhlenbeck (OU) process with
\begin{equation}
\mathbb E_\xi[\xi(t)]=0,~~\mathbb E_\xi[\xi(t)\xi(s)]=\sigma_\xi^2e^{-|t-s|/\tau_c},
\label{eq:spin_ou_correlation}
\end{equation}
where $\sigma_\xi$ and $\tau_c$ denote the standard deviation and correlation time, respectively. In the sensing process, the initial state is fixed to $|+x\rangle$ and the QFI is calculated at $\theta_0=0$. The control fields are optimized using the algorithm for stochastic unitary evolution described in Sec.~\ref{subsec:Generalization-SHE}.

Before studying AC-field sensing under colored noise, we first consider the noise-free case to gain insight into the sensing dynamics. In the absence of a control amplitude bound, one can prove that ideal $\pi_x$-pulse applied at the signal zero crossings, i.e.,  $\omega t_k= (k-\frac12)\pi$ with $k=1, 2, \cdots$ (CPMG pulses), yields the optimal QFI 
\begin{equation}
    F_\theta (T)= \left[\int^T_0 dt |\cos \omega t|\right]^2.
\end{equation}
The corresponding control field reads
$u_x(t)=\pi\sum_{t_k<T}\delta(t-t_k)$. These equally spaced pulses form a CPMG sequence when $\omega T=n\pi$. Specifically, the Bloch vector $\bm{r}(t)$ and its derivative $\bm{s}(t)$ in this case read $\bm{r}(t)=(1,0,0)$ and $\bm{s}=(0, \int^{t}_{0} d\tau |\cos \omega \tau|, 0)$, yielding the sensitivity-tangent speed, the alignment factor, and the accumulation phase
\begin{equation*}
    J_F= \frac{1}{2}|\cos \omega t|,~~ \kappa= 1, ~~ \beta=0.
\end{equation*}
Under the amplitude constraint, ideal pulses are replaced by rectangular pulses centered at $t_k$, with amplitude $E$  and duration $\Delta t= \pi/E$. The PMP-based
numerical optimization recovers a protocol closely matching this pulse sequence, as shown in Fig.~\ref{fig:spin_ac} (a1--a4). The resulting QFI dynamics and the evolution of $J_F \kappa$ and $\beta$ closely resemble those of the ideal case.  In particular, the sensitivity growth rate $d\sqrt{F_\theta}/dt=2J_F\kappa\cos\beta$ shows that ideal $\pi_x$ pulses applied at the signal zero crossings transform the cosine oscillation into a full-wave rectified cosine waveform, yielding monotonically increasing QFI.

At a fixed interrogation time, the results for weak and strong noises are shown in Fig.~\ref{fig:spin_ac} (b1--b4) and (c1--c4), respectively. Here, we set $\omega=1.0$, $\omega \tau_c=\pi$, and fix $\omega T=3\pi$. The noise causes dephasing of the sensor. Without control, the Bloch vector $\bm{r}(t)$ remains along the $x$-direction with $r_y=r_z=0$ and 
\begin{equation}
    r_x(T)=e^{-\sigma_\xi^2\tau_c\left[T-\tau_c \left(1-e^{-T/\tau_c}\right)\right]}.
\end{equation}
Consequently, for $T\gg\tau_c$, the QFI oscillates within an exponentially decaying envelope,
\begin{equation}
F_\theta(T)\approx \frac{\sin^2 \omega T}{\omega^2}e^{-2\sigma_\xi^2\tau_c T}.
\label{eq:spin_ac_free_qfi}
\end{equation}
Applying $\pi_x$ pulses promotes the accumulation of the AC signal and suppresses dephasing, mitigating the decay of the QFI.  For ideal pulses at $\omega T=n\pi$, we have 
\begin{equation}
  r_x(T)=e^{-\Gamma_n},~~ F_\theta= 4\frac{n^2}{\omega^2} e^{- 2\Gamma_n},
\end{equation}
where the dephasing exponent is
\begin{equation}
    \Gamma_n=\sigma^2_\xi\tau_c[\eta T-\tau_c (1-{\rm sech} q)^2 (1-(-e^{-2q})^{n})],
\end{equation}
with $q=\pi/2\omega \tau_c$ and $\eta=1-\frac{1}{q}\tanh q$. This expression reveals that the dephasing rate is suppressed by a factor $\eta$ and predicts a QFI peak near 
\begin{equation}\label{eq:nmax}
n_{\rm max}\approx [\eta \pi \sigma^2_\xi \tau_c/\omega]^{-1},   
\end{equation}
larger than which $F_{\theta}$ decreases with $n$. 
                
At a fixed $T$, we maximize the terminal QFI under the control-amplitude constraint using the projected gradient ascent method in Sec. \ref{projected gradient ascent}. The gradients are evaluated by the discrete adjoint method for noisy dynamics and analytically in the absence of noise. For the single-axis control considered here, the projection enforces $|u_x(t)|\le E$, while $u_y(t)=0$. We optimize six initial controls selected from a pool of randomly generated waveforms and further optimize the best solution on finer time grids. For $\omega T=3\pi$, the initial grid contains 144 equal time intervals. The final grids contain 288, 288, and 1152 intervals for $\sigma_\xi/\omega=0$, $0.4$, and $0.8$, respectively. For each noisy case, the objective function and its gradient are evaluated using $10240$ independent realizations of a stationary Ornstein--Uhlenbeck process, held fixed throughout the optimization. After optimization, the QFI of the ensemble-averaged density matrix is evaluated using a separate set of $10240$ noise realizations, independent of those used for optimization. The same evaluation realizations are used for the optimized-control, CPMG, and uncontrolled protocols.

The numerical result shown in Fig.~\ref{fig:spin_ac} (b1--b4) support the above analysis for finite-width pulses. Under non-ideal CPMG, $F_\theta$ increases almost monotonically under CPMG pulses with $n_{\rm max}\approx 7$.  Furthermore, $r_x(t)$ also confirms that dephasing is strongly suppressed. For the parameters studied, our numerical searches found no appreciable improvement from allowing a nonzero \(u_y(t)\). We therefore focus on single-axis control. We therefore restrict the optimization to single-axis control by setting $u_y(t)=0$ throughout this section. For weak noise, the optimized waveform shown in Fig.~\ref{fig:spin_ac} (b2) closely resembles the CMPG plus sequence, with small changes in pulse shape providing only a slight increase in the final QFI. This means that the CPMG pulse is nearly optimal for AC sensing under weak noise. Physically, this is because CPMG pulses already suppress dephasing effectively. Further control offers little benefit, as the small gain in coherence improvement is offset by reducing the AC signal accumulation. The sensitivity growth rate in panel (b4) exhibits modest deviations from the rectified cosine waveform of the noise-free case.

\begin{figure}[thb]
    \centering
    \includegraphics[width=0.95\linewidth]{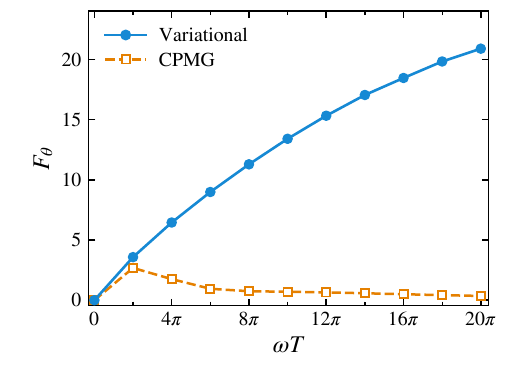}
    \caption{Terminal QFI at different interrogation times for AC-field sensing under colored noise. Filled blue circles show the results obtained from optimized control, and open orange squares denotes the results obtained from finite-duration CPMG control. Parameters are $\sigma_\xi/\omega=0.8$, $\omega\tau_c=\pi$, $E/\omega=8$, and $u_y=0$, with initial state $|+x\rangle$ and operating point $\theta_0=0$.}
\label{fig:terminal-qfi}
\end{figure}

Under stronger noise, the same analysis predicts an earlier turnover of the QFI, as dephasing increasingly offsets signal accumulation. The results for $\sigma_\xi/\omega=0.8$ are shown in Fig.~\ref{fig:spin_ac} (c1--c4). Here, substantial dephasing persists even under CPMG control, as shown in (c3). The accumulated AC signal is totally destroyed by the dephasing effect, causing a QFI that no longer increases shown in (c1) and $c(4)$ with $n_{\rm max} \approx 2$. Numerical optimization yields a very complex control design, shown in panel (c2). The evolution of $r_x(t)$ reveals that the optimized control preserves coherence more effectively than the CPMG sequence. The similar oscillatory behavior of $r_x(t)$ also suggests that the optimized protocol retains features of the CPMG dynamics while preserving coherence more effectively. In addition, the sensitivity growth rate in panel (c4) further shows that the optimized protocol replaces much of the negative contribution under CPMG control with positive growth, partially restoring
a rectified cosine-like waveform. Together with the enhanced QFI, these observations suggest that the optimized control suppresses dephasing without substantially compromising AC-signal accumulation, allowing the QFI to continue increasing with time.

These numerical results demonstrate the benefit of optimized control for AC-field sensing under strong noise.  Compared with CPMG pulses, the optimized waveform suppresses decoherence more effectively without substantially compromising AC-signal accumulation, thereby enhancing the final QFI.  The complex waveform highlights the nontrivial nature of control design and illustrates the effectiveness of the PMP-based method in identifying improved sensing protocols.

For $\sigma_\xi=0.8\omega$, Fig.~\ref{fig:terminal-qfi} compares the terminal QFI obtained with controls optimized separately at each interrogation time $T$ with that obtained using CPMG pulses. Here, the OU noise average is evaluated deterministically by solving the stochastic Liouville equation in a Hermite basis~\cite{Kubo1963,Tanimura2006}. The hierarchy and its parameter derivative are propagated together, and their zeroth components give the ensemble-averaged density matrix and its derivative for evaluating the QFI. This retains the finite noise correlation time, and convergence is checked by increasing the hierarchy truncation order. Among the sampled interrogation times, the QFI under CPMG reaches its maximum at $\omega T=2\pi$, consistent with the ideal-pulse estimate of $n_{\rm max}$. In contrast, the optimized terminal QFI increases across the sampled interrogation times up to $\omega T=20\pi$. This growth persists at interrogation times far beyond the uncontrolled coherence time $T_2^\ast$, for which $\omega T_2^\ast\approx2.0$, demonstrating the advantage of optimized control for AC-field sensing under strong noise.

\section{Conclusions and discussions}\label{sec:Conclusions}
We investigated optimal quantum metrology for a fixed initial probe state under bounded, restricted Hamiltonian controls.  Building on Pontryagin's maximum principle, we formulated the optimization using augmented state--costate equations and control gradients for multiple control channels subject to a weighted amplitude constraint. The formulation accommodates stochastic Hamiltonian evolution and time-local master equations. We also introduced a geometric description of sensitivity accumulation that separates the sensitivity-tangent speed, alignment factor, and accumulation phase, providing a means of interpreting the optimized protocols.

Two spin-$1/2$ examples illustrate the resulting sensing enhancement. For transverse-field sensing, time-dependent control outperforms optimized constant control at long interrogation times. The state and sensitivity trajectories reveal an initial
steering stage that sacrifices immediate sensitivity growth to establish a more favorable trajectory for subsequent encoding, followed by a final adjustment that further increases the terminal QFI. For AC-field sensing under colored noise, optimized controls
closely resemble CPMG pulses in the weak-noise regime, whereas more structured waveforms provide a substantial advantage under strong noise by improving the balance between coherence preservation and sensitivity accumulation.

These results show that nontrivial bounded controls can improve the use of a prescribed probe even for a single spin. Future work could investigate the robustness of these protocols to control errors and uncertainty in the noise model, incorporate bandwidth
constraints, and extend the analysis to interacting many-body sensors. In realistic many-body sensors, probe preparation, Hamiltonian control, and readout are all subject to experimental restrictions~\cite{Dooley2016,YangEtAl2022}. In particular, readout may be limited to simple local or collective observables and finite detection resolution~\cite{Davis2016Readout,Hosten2016Magnification,Nolan2017Readout, Len2022}. The present variational framework can be extended to search for controls that maximize the sensitivity of a prescribed many-body probe within these constraints. 

\bibliography{refs}
\end{document}